\documentstyle[prb,aps,twocolumn,amssymb,amsfonts,epsfig]{revtex}

\begin{document}


\title{Anti-ferromagnetism, spin-phonon interaction and the local-density approximation
 in high-T$_C$ superconductors.}

\author
{T. Jarlborg}

\address{D\'epartement de Physique de la Mati\`ere Condens\'ee,
Universit\'e de Gen\`eve, 24 Quai Ernest Ansermet, CH-1211 Gen\`eve 4,
Switzerland} 

\date{\today}
\maketitle

\begin{abstract}
Results from different sets of band calculations for undoped and doped HgBa$_2$CuO$_4$
show that small changes in localization can lead to very different ground states.
 The normal LDA results are compared with 'modified' LDA results, in which different linearization
energies make the O-p band more localized. The ground states in the normal calculations are far from
the anti-ferromagnetic ones, while nearly AFM states are found in the modified calculations.
The proximity of an AFM state in the doped system leads to increased 
$\lambda_{sf}$, and the modified band structure has favorable
conditions for spin-phonon coupling and superconductivity mediated by spin fluctuations. 


\end{abstract}


The Fermi surfaces (FS) of high-T$_C$ superconductors contain
barrel-like cylinders, which originate from the 2-dimensional CuO-planes. Other smaller
FS-pockets or ridges are found in the more complex high-T$_C$ materials, but it is
probable that superconductivity is localized to the barrels, since they are common to all of the
high-$T_C$ cuprates. Electronic structure calculations, based on the local density approximation, 
LDA \cite{lda},
describe the FS correctly in the doped cuprates. But
the anti-ferromagnetic (AFM), insulating ground state
of many undoped materials is absent \cite{san}. 
Thus, the band structure of the doped, metallic and superconducting
system can be described within LDA, while the competing AFM insulating state 
in the undoped case is too
high in energy to enter into the picture of possible ground states. This is a major
problem when using LDA (and gradient corrected density functionals, GGA \cite{gga}) for many oxides, and
the consequence is that most theories for high-$T_C$ superconductivity are oriented towards models
of the band structure. This is generally done
through introduction of parameters in so-called $t$-$J$-models, or through models
for strong on-site correlation.

It is not satisfactory to abandon the density functional (DF) theory
without knowing the reason for this partial failure, especially since it is
working so well for the large majority of other materials. Correlation is often cited as a  
possible reason, but why should this suddenly become large because of a slight
change in chemical potential via different doping? The energies involved in
on-site correlation, $U$, are large, several volts are typical, so $U$ has to vary
extremely fast with doping.
 Here I suggest
another possibility; that rather small changes of the potential can make that
two very different configurations have similar total energies and enter
into competition. Thus, the situation could be like the one for the ground state
of iron, where LDA calculations predict nonmagnetic fcc as the ground
state, while GGA correctly predicts the FM bcc state as the stable one \cite{bmj}.
The differences in total energy between the different configurations are small
within each DF-potential, only the fine details make one state more stable than the
other. Still, the changes in potential given by GGA are small, and
many other properties of Fe can be calculated correctly by the use of
both DF-potentials. The GGA applied to oxides make no large improvement compared to LDA.

Band calculations, using the linear Muffin-Tin Orbital method (LMTO) \cite{lmto,ja} and LDA for 
one of the high-T$_C$ materials, HgBa$_2$CuO$_4$ (HBCO),
showed that spin fluctuations are coupled to phonons within the CuO plane \cite{tjb}.
The qualitative results for the doped case are consistent with many observations in high-T$_C$ materials such
as the existence of stripes, pseudogaps in the density-of-states (DOS), 
isotope shifts, AFM fluctuations, phonon
softening in the doped materials etc.. But the amplitude of the coupling is too low for a high T$_C$.
The likely problem with LDA is that the
total energy of the AFM ground state is much above the nonmagnetic state in undoped HBCO.
We shall analyze the ingredients of the total energy of an AFM configuration by comparing it
to the case with no spin-polarization. The exchange energy is always in favor of spin-polarization,
as is well known from the Stoner model for FM \cite{jf}. In contrast to the FM case,
also the kinetic energy favors
magnetism if the AFM state will open a gap or pseudogap at the Fermi energy, $E_F$, as was discussed
previously for HBCO \cite{tjrap}. But the AFM state, where the potential for one spin
show a spacial modulation, costs hybridization energy. The tail of a wavefunction of one spin from 
one site enters another site where the spin is different, making the
conditions for hybridization different from the unpolarized case. 
The FM case is well described by a rigid band
shift of the entire majority band structure relative to the minority one, but in the AFM
case there is no such shift.
One can expect that the hybridization would decrease
for more localized states, since the tails become smaller. The plane oxygens
(between Cu sites of opposite spins) have their p-states hybridizing with the d-states
on Cu, and vice-versa. The idea is to make the O-p states more localized
so the interaction between Cu of opposite spins can decrease and thereby
 minimize the cost of hybridization energy for an AFM state.

The logarithmic derivatives, $D_{\ell}$, of the LMTO method \cite{lmto}, provide information about
localization of each band $\ell$. The values of $D_{\ell}$ are negative in the band region, and they decrease
continuously for increasing energy until they become infinite at the top of the $\ell$-band.
The localization is best when the energy of the state is at the top of the band. 
The LDA-value of $D_{\ell}$ for the oxygen p-band in HBCO is about -5 at $E_F$, showing that the band
is not completely filled. The hybridization energy should diminish when $D_{\ell} \rightarrow -\infty$,  
corresponding to a very localized state at $E_F$. 
States far below $E_F$ are extended,  but such states are less affected by the spin splitted potential.
In figure 1 it is seen that the nonmagnetic and AFM DOS functions differ only in the
upper part of the Cu-d band.  
This is because the radial extent of the exchange splitting of the potential
is localized within the Cu-atom, and the splitting of an extended state is
relatively small.

The optimal choice of the linearization energies for the $\ell$-bands in LMTO is at the center of each band.
One way of making the states more localized is to
do the linearization at lower energies, at the bottom of the bands. 
This makes the self-consistent $D_{\ell}$-value at $E_F$
positive, showing that the top of the O p-band goes down compared to the normal case.
This is the main difference for making modified
(more localized) bands without changing the LDA-potential. It is not a 'good' way of calculating
the LDA bandstructure, but as will be shown, the results serve as examples of how to generate 
an AFM, insulating
state in the undoped case 
and stronger AFM fluctuations in the doped case.
A full gap appears in the DOS of the undoped HBCO if a staggered field of $\pm 23 mRy$ is applied
within the Cu sites in the normal LDA calculations.  Calculations with 
lower choices of linearization energies ('modified-I')
give gaps already at smaller applied fields, $\sim$10 mRy,
see figure 1. 
Stronger changes are found when an additional modification ('modified-II') is made in the calculation of the
Madelung potential. The method for calculating the Madelung potential uses
non-overlaping MT-spheres in the calculation
of the point charges, that are smaller than the Wigner-Seitz (WS) spheres 
used in the LMTO scattering matrix. 
An interstitial volume between the MT and WS spheres has an interstitial charge. This procedure
has a moderating effect on the Madelung potential for compounds \cite{ja}. An additional
lowering of the p-band is observed when using oxygen MT- and WS-spheres of equal size.
For this 'modified-II' LDA calculation it is
just possible to obtain a gap and AFM without an applied field, see figure 2. The moment is 0.23 $\mu_B$
per Cu atom.

The band widths of the d- and p-bands from the CuO-plane 
are reduced by 0.02 Ry in the 'modified-II' LDA calculation. A more important
change is that the top of the O-p band goes down from 0.14 Ry above $E_F$ to about 0.17 Ry
below $E_F$. The top of the Cu-d band goes up relative to $E_F$, by 0.03 Ry. This 
can be seen in the DOS, in which the shoulder of large DOS moves upwards closer to $E_F$. The
same happens when large fields are applied and the shoulder can even reach the position
of $E_F$ for the doped case, cf figure 1. 
This is an extreme, not realistic case and it will hardly help the AFM state.  

The charge distribution is changed in the modified case. As can be expected from the changes 
in band positions there is an increase of the 
the O p-charge by 0.11 e$^-$, while the Cu d-charge decreases by 0.09 e$^-$ per atom.
These changes follow the same trends that were found from calculations
with improved correlation added to the LDA
for the CuO planes in the 'infinite-layer' structure  \cite{stoll}. 
No detailed comparison
with results from the method of Stollhoff is possible because of
different basis, different structure, and so on.  But the largest charge redistribution
caused by correlation \cite{stoll}, a decrease of the $x^2-y^2$ Cu d-charge by 0.38 $e^-$ and an increase
of the O p-charge along the bonding by 0.19 $e^-$, are in line with the present results.
The result of Stollhoff call for moderate corrections of LDA due to correlation, but
it also excludes the limit of strong correlation and the Mott-Hubbard model \cite{stoll}.  

Before turning to the doped system it might be appropriate to make some 
comments about the energy scale of the perturbations.
The energy of the magnetic field, $\mu H$=23 mRy, required to make the undoped system an AFM insulator
in LDA calculations, seems large in view of the fact that 1 mRy corresponds to a magnetic field
of 230 T. However, this is not much on the scale of other electronic energies in the band calculation.
Total energies calculated in LDA and GGA can differ by a few Ry per atom, although differences in
total energies between different configurations (that are to be compared with physical properties)
and other main features of the band results
are reasonable from both DF potentials. 
Thus, the value 23 mRy can by itself not be taken as an indication that LDA
is very far from a good description of the ground state. The fact that $\mu H$ can approach 0 mRy and produce
an AFM insulating state in 'modified' LDA calculations, shows that electronic energies
associated with charge localization can easily replace the energy coming from an applied magnetic field.
Moreover, it can be recalled that when LDA bands are replaced by band models including strong correlation
one needs to apply larger energies in terms of on-site correlation $U$, 
which typically are of the order of 5 eV ($\sim$400 mRy).  

The next step is to do the same modification for the hole doped case. 
 The pseudogap will appear at the new position of $E_F$ if
the AFM state is modulated, as has been shown from calculations using 
the virtual crystal approximation for a supercell oriented along 
the CuO bond direction \cite{tjrap}. 
The wavelengths
of spinwaves and phonons differ by a factor of two for an optimal spin-phonon coupling,
and they depend on the doping.
This, and other features of the spin-phonon coupling can be found in previous publications 
\cite{tjrap,tjb}, but one may note that the relatively short supercell considered here 
corresponds to phonon and spin waves in an overdoped system. 
The pseudogap, the large exchange enhancements and 
magnetic moments on the Cu-sites, are all indications of 
a metal-insulator transition.
The normal LDA exchange enhancement, $S$,
evaluated as the exchange splitting of Cu-d band divided by the applied field, 
is about 2.0
when the cell contains the phonon distortion and the spinwave.  The enhancements in the 'modified'
LDA calculation are much larger, see Table 1. The moments ($m$) 
are larger than in the normal LDA,  
and they are normally proportional to the applied field, $\mu H$, $m=M \cdot{\mu H}$.

The calculation of $\lambda_{sf}$
is made assuming harmonic conditions for a 'frozen' spinwave \cite{tjb,tjfe}:
\begin{equation}
\label{eq:lamsf}
\lambda_{sf} = N I^2/2F 
\end{equation}
where $N$ is the paramagnetic DOS at E$_F$ (100-115 states/cell). The matrix element, $I$, is the derivative
of the potential with
respect to $m$. In the harmonic limit; $ < \delta V > = I \cdot m$, and $I$ is calculated 
from the FS average of the difference of band energies, $\epsilon^m_k - \epsilon^0_k$,
between the cases with and without magnetization. It is assumed that
the difference in total energy between the configuration with moment $m$ and at zero moment,
$E_m - E_0$, depends almost quadratically on $m$, $E_m - E_0 = F m^2$, similar to the dependence
of the total energy on the atomic displacements, $\Delta r$, for harmonic phonons.

Table I shows the results of $N$, $S$, $M$, $I^2$, $F$ and $\lambda_{sf}$ compared to the results from
the ordinary LDA calculations, case (b) in the previous work \cite{tjb}. 
The enhancement $S$ and the factor $M$ show qualitatively how the system gets
closer to an AFM transition. 
The calculations are made for applied fields up to 50 mRy in the ordinary LDA, while up to 30 mRy
in the modified cases. The moments are larger in the modified calculations, when the AFM
instability is closer and the harmonic
behavior is less evident.
The proximity of the AFM instability makes the ingredients ($I^2$ and $F$)
to $\lambda_{sf}$ smaller, and the  
relative errors in $F$ and $\lambda_{sf}$ are larger. It is difficult to calculate
$\lambda_{sf}$ in the limit $m \rightarrow 0$, since an estimation of $F$ from
the differences of total energies between small and zero fields is too uncertain.
Using the high or low $m$-limit for $\lambda_{sf}$
makes no big difference in the ordinary LDA, but in the modified calculations it is the low-$m$ limit 
that gives the largest
$\lambda_{sf}$. The results displayed in
Table I  are based on results from
larger $m$. Despite the increase in relative errors it seems clear that the
coupling constant for spin-fluctuations is larger
in the modified LDA calculations, and it can be large enough for a high T$_C$. 
However, the calculation of T$_C$ requires further considerations, whether the prefactor
in a BCS-like formula should be given by the excitation energy from phonons or from spin waves, whether the
corrections due to non-harmonicity are important, etc..
No theory is yet known to describe superconductivity within a spin-phonon coupled system, 
but it is reasonable to assume that a larger $\lambda_{sf}$ should give a larger T$_C$.

The main effect from the modified band calculation is that the potential becomes more attractive 
within the oxygens compared
to the normal LDA. The HBCO structure contains Ba, Hg and apical oxygens, and the influence from
these sites on the potential within the CuO plane gives additional
possibilities for AFM.  For instance, it was found that the relative
positions between the CuO plane and Ba or apical oxygens are important for the local
exchange enhancement on Cu \cite{tjrap,tjb}. It remains to be investigated in detail
how real phonons with longer wavelengths interact with spin waves in underdoped systems. 
Various reasons can be suggested to lead to a more attractive potential on oxygen sites,
such as non-locality, kinetic energy
or correlation, but this will not be discussed here.  

In conclusion, two band calculations using the same LDA potential, but different
linearizations, show very different results for the stability of the AFM state of undoped HBCO
compared to ordinary LDA.
The same exercise for doped HBCO show that the modified LDA will enhance 
the spin-phonon coupling and $\lambda_{sf}$.
It is argued that LDA might need rather modest corrections in order to promote an AFM, insulating
state in oxides. The resulting charge transfer is in agreement with an independent investigation 
of moderate correlation \cite{stoll}. The gap is in this case an ordinary band gap, caused by
modulations of AFM within the CuO plane. This is
in contrast to the Mott-Hubbard gap due to very strong correlation.


\newpage

\begin{table}
\caption{Calculated parameters (see text) for ordinary and modified LDA for a spinwave co-existing
with a halfbreathing phonon as in ref. \cite{tjb}. The units are for $N$
states/cell/Ry, for $M$: $\mu_B/mRy/cell$, for 
$I^2$: $10^{-3}Ry^2/\mu_B^2$, and for $F$: $mRy/cell/\mu_B^2$.
}
\begin{tabular}{ccccccc}     
  & $N$ & $S$ & $M$ & $I^2$ & $F$ & $\lambda_{sf}$    \\
\tableline
ordinary LDA & 100 & 2.0 & 0.05 & 0.12 & 33 & 0.2 \\
modified-I LDA & 110 & 5.0 & 0.15 & 0.03 & 4 & 0.4 \\
modified-II LDA & 115 & 6.5 & 0.19 & 0.03 & 3 & 0.5 \\
\end{tabular}
\end{table}

\begin{figure}[tb!]

\leavevmode\begin{center}\epsfxsize8.6cm\epsfbox{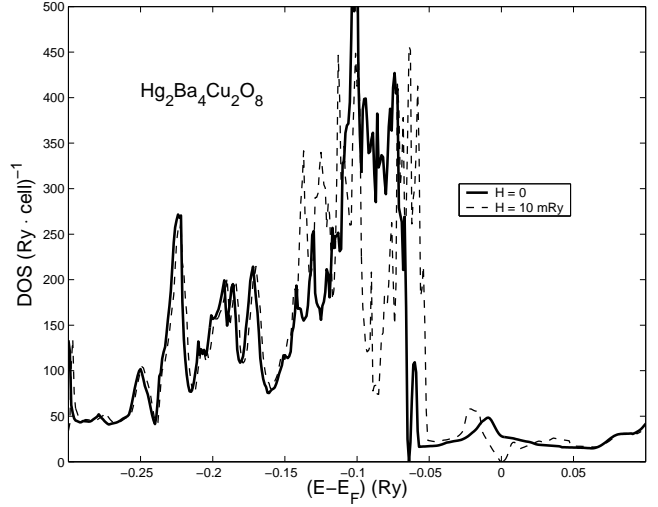}\end{center}
\caption{Calculated DOS of Hg$_2$Ba$_4$Cu$_2$O$_8$ in which the linearization energies
are low compared to the band centers ('modified-I', see text).  A staggered field of $\pm 10 mRy$
on the Cu makes a gap at $E_F$. Note that the AFM and the non-polarized DOS differ
only at the upper part of the Cu-d band. 
}
\end{figure}
\begin{figure}[tb!]
\leavevmode\begin{center}\epsfxsize8.6cm\epsfbox{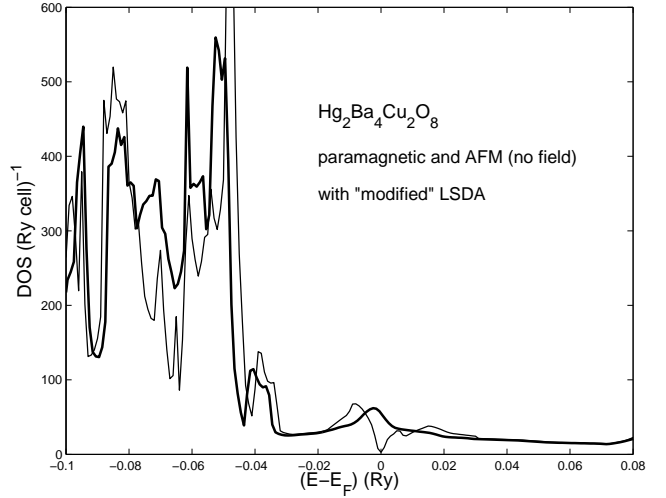}\end{center}
\caption{Paramagnetic (bold line) and AFM (thin line) DOS near $E_F$ of Hg$_2$Ba$_4$Cu$_2$O$_8$ calculated
for the lowest choice of linearization energies and a modified Madelung charge ('modified-II', see text). 
}
\end{figure}

\end{document}